\documentclass[aps,pre,reprint,superscriptaddress,amsmath,amssymb,showpacs,floatfix]{revtex4-1}
\usepackage{graphicx}
\usepackage{bm}
\usepackage{color}
\usepackage{soul}

\begin{document}
\title{Finite-time quantum Otto engine: Surpassing the quasi-static efficiency due to friction}
\author{Sangyun Lee}
\affiliation{Department of Physics, Korea Advanced Institute of Science and Technology, Daejeon 34051, Korea }

\author{Meesoon Ha}
\email[]{msha@chosun.ac.kr}
\affiliation{Department of Physics Education, Chosun University, Gwangju 61452, Korea}

\author{Jong-Min Park}
\affiliation{School of Physics, Korea Institute for Advanced Study, Seoul, 02455, Korea}

\author{Hawoong Jeong}
\email[]{hjeong@kaist.edu}
\affiliation{Department of Physics and Institute for the BioCentury, Korea Advanced Institute of Science and Technology, Daejeon 34141, Korea}

\date{\today}

\begin{abstract}
In finite-time quantum heat engines, some work is consumed to drive a working fluid accompanying coherence, which is called `friction'. To understand the role of friction in quantum thermodynamics, we present a couple of finite-time quantum Otto cycles with two different baths:  Agarwal versus Lindbladian. We solve them exactly and compare the performance of the Agarwal engine with that of the Lindbladian engine. In particular, we find remarkable and counterintuitive results that the performance of the Agarwal engine due to friction can be much higher than that in the quasistatic limit with the Otto efficiency, and the power of the Lindbladian engine can be nonzero in the short-time limit. Based on additional numerical calculations of these outcomes, we discuss possible origins of such differences between two engines and reveal them. Our results imply that even with an equilibrium bath, a nonequilibrium working fluid brings on the higher performance than what an equilibrium working fluid does. 
\end{abstract}

\maketitle
 
\section{Introduction} 
\label{intro}

How quantumness plays a role in thermodynamics is one of the interesting and important questions to understand quantum phenomena, which is the so-called {\em quantum thermodynamics}~\cite{Gemmer2009QT} that concerns the relation between quantum mechanics and thermodynamics. In a sense, to study quantum heat engines, Ref.~\cite{Kosloff2014-Review} (and references therein) has provided useful frameworks for further theoretical and experimental developments. 

A quantum heat engine is a cycle with thermodynamic processes, and its working fluid is a quantum system with coherence, entanglement, and discrete energy levels. Due to the development of experimental techniques, it has been realized in various ways~\cite{Rossnagel2016single,josefsson2018quantum,deAssis2019PRL-QOHE-experiment, Peterson2019Experimental}, and various heat baths have also been considered:  a coherent bath was used to exceed the Carnot efficiency, and decoherent one was introduced to find the signature of quantumness~\cite{Uzdin2015Equivalence, Camati2019Coherence, Guff2019Coherent}.
Squeezed bath~\cite{Robnagel2014Nanoscale} also allowed the efficiency to be beyond the Carnot efficiency due to the nonequilibrium resource. Moreover, it is known that a quantum phase transition can be used to increase the efficiency~\cite{Ma2017quantum} or decrease it~\cite{Kloc2019PRE-FTQOC}. 


Owing to the discovery of the trade-off relation between the power and the efficiency of the engine~\cite{Shiraishi2016trade-off, Pietzonka2018Universal, Ma2018Universal} as well as the development of the shortcut-to-adiabaticity technique~\cite{Abah2019PRE-STA}, the finite-time quantum heat engine has steadily gathered significant attention, where the working fluid can have coherence without any special bath, such as a squeezed or coherent bath. When Hamiltonians at different times do not commute, a portion of work is used to generate coherence. At last, it is dissipated when the system is coupled to a heat bath later. 

Such a mechanism is regarded as a quantum analog of {\em friction}.
There have been many ways to measure friction in quantum heat engines~\cite{Feldmann2003Quantum,Rezek2006Irreversible,Chen2010Fast, Alecce2015Quantum}, but we focus only on the friction by the power term that is required to drive the working fluid in the finite-time mode, which has been in Otto heat engines~\cite{Feldmann2003Quantum,Rezek2006Irreversible}.
%

\begin{figure}[b]
 \includegraphics*[width=0.8\columnwidth]{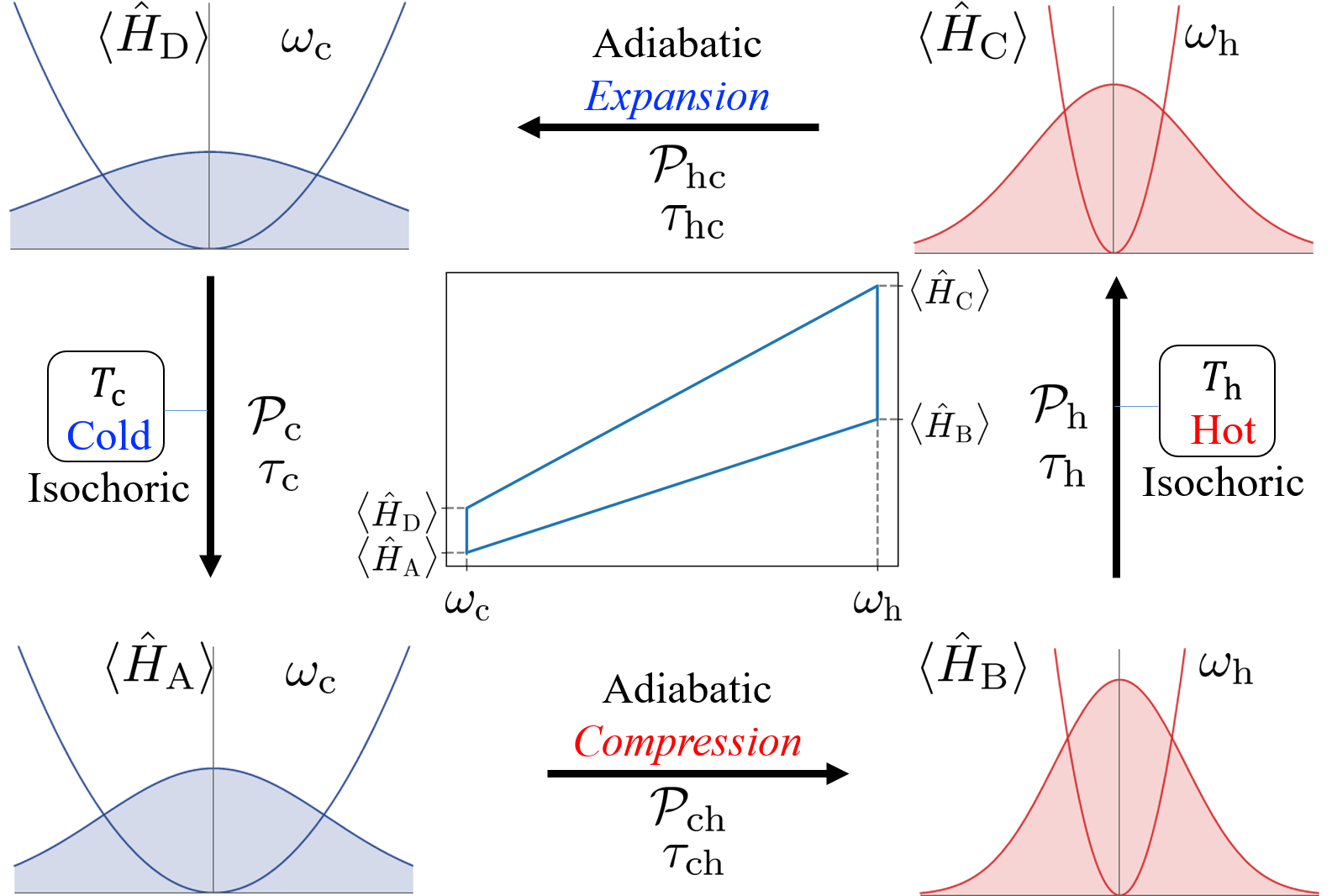}
\caption{ A finite-time quantum Otto cycle is schematically illustrated with harmonic potentials and Wigner functions, which consists of isochoric and adiabatic processes. In the isochore, the working fluid exchanges heat with heat bath of temperature $T_{\rm h}$ ($T_{\rm c}$) by the propagator, $\mathcal{P}_{\rm h}$ ($\mathcal{P}_{\rm c})$,  of the vector, $(\langle\hat{H}\rangle, \langle\hat{L}\rangle, \langle\hat{D}\rangle, \langle\hat{I}\rangle)$ for the process time 
$\tau_{\rm h}$ ($\tau_{\rm c}$), whereas, in the adiabatic expansion (compression), the internal energy change of the working fluid becomes work by $\mathcal{P}_{\rm hc}$ ($\mathcal{P}_{\rm ch}$) for $\tau_{\rm hc}$ ($\tau_{\rm ch}$). The total-energy expectation of the working fluid is drawn as a function of $\omega$ in the middle panel.}
\label{fig:conceptual}
\end{figure}

The Otto cycle (see Fig.~\ref{fig:conceptual}) has been widely studied due to its analytic tractability~\cite{Feldmann2003Quantum,Rezek2006Irreversible,Zheng2014Work, Abah2012single,Abah2014Efficiency,*Abah2016EPL-Otto-friger, Niedenzu2016On,Ma2018Optimal, Chen2019Boosting,*Chen2019Achieve,JM2019Quantum,Kloc2019PRE-FTQOC,Abah2019PRE-STA}. It has been reported that the quantum Otto engine can also be used as a precise thermometer~\cite{Hofer2017Quantum} and the Otto engine with the finite power and the quasistatic efficiency can be achieved by the shortcut-to-adiabaticity technique~\cite{Abah2019PRE-STA}. Most recently, it has also been realized with the nuclear magnetic resonance spectrometer~\cite{deAssis2019PRL-QOHE-experiment, Peterson2019Experimental} and its quasistatic efficiency has been beaten in the finite-time mode with a heat bath of effective negative temperatures~\cite{deAssis2019PRL-QOHE-experiment}. 

In this paper, we consider two quantum Otto cycles in finite-time frameworks with a time-dependent harmonic oscillator, exactly solve their performances, and discuss the role of friction in them as a quantum effect. To describe a quantum system connecting to the corresponding heat bath, we revisit the well-known Lindblad master equation (Lindblad bath, {\bf L})~\cite{Insinga2016Thermodynamical} with a propagator of a dynamical semigroup, and the Agarwal master equation (Agarwal bath, {\bf A}) ~\cite{Agarwal1970PRA,*Agarwal1971Brownian,Agarwal2013quantum} (and references therein) as paradigmatic models. In particular, we focus on how they are different from each other in the finite-time mode. Finally, it turns out that they exhibit fruitful physics with remarkable and counterintuitive results. 

In the quasistatic limit, regardless of detailed model approaches, the Otto efficiency is only determined by the volume ratio between hot and cold isochores. As the cycle time becomes infinity, its power becomes eventually zero. As a result, the quasistatic limiting performance of quantum Otto heat engines is rather trivial, so that two baths (propagators) do not make any difference between their performances in the quasistatic limit. However, in the finite-time mode, they can be different due to the role of friction and the setup of heat bath. To the best of our knowledge, the case of the Lindblad bath was exactly solved, but the Agarwal case has not been exactly solved yet.  So, in this paper, we exactly solve the Agarwal case and compare it with the Lindbladian and discuss the origin of counterintuitive results.

Both the Agarwal bath and the Lindbladian bath are based on the Born-Markov approximation~\cite{Massignan2015QuantumBrown}. 
For the Lindbladian case, the secular approximation is applied as neglecting rapidly oscillating terms~\cite{Breuer2006TheTheory}.  While the Lindbladian master equation is appropriate to model Otto engines based on quantum optics and the completely positive map, the Agarwal master equation corresponds to the Born-Markov master equation without the secular approximation and the positive map. It is known that by adding some terms a master equation of a positive map can be completely positive and with the fixed master equation does not satisfy translational invariance of dissipation and detailed balance simultaneously~\cite{Diosi1993Caldeira, Alicki2018}. 
A possible experiment has been proposed with an impurity in a quantum gas to observe the system~\cite{Massignan2015QuantumBrown}.  Although our Agarwal engine might be too simple to describe real experiments, we believe that this research paves the way to understand the differences between an engine with quantum optics and an engine with an impurity in an ultracold quantum gas.

The rest of this paper is organized as follows: In Sec.~\ref{system}, we describe a finite-time quantum Otto heat engine with the harmonic oscillator and present two different types of baths, where the performances (efficiency and power) of Otto cycles are denoted with the analytic forms of work and heat. In Sec.~\ref{result}, we exactly solve the performance of each case as well as numerical enumerations for related physical quantities, where we argue the possible origins of the differences between two cases and confirm them. In particular, we focus on the performance in the short-time limit and near resonance conditions, where it gets better counterintuitively. Finally, in Sec.~\ref{conclusion}, we conclude this paper with a summary and some remarks.

\section{System}
\label{system}

\subsection{Otto cycle}

As illustrated in Fig.~\ref{fig:conceptual}, an Otto cycle consists of two isochoric (constant volume) and two adiabatic (no heat transfer) processes. In the isochore, there is no external force and no explicit time dependence on the Hamiltonian, $\hat{H}(t)$, so that all the energy change of the engine becomes heat.
We consider a couple of heat baths for the isochores, which drive a system into the same equilibrium state, the Lindblad bath versus the Agarwal bath.

The governing equation of the density matrix  is 
\begin{align}
    \frac{{\rm d} \hat{ \rho } (t) }{{\rm d} t} = -\frac{i}{\hbar}[ \hat H(t), \hat\rho(t)] +\mathcal{L}_{k}(\hat{\rho}(t)).
    \label{eq:general}
\end{align}
where ${\mathcal L}_{k}$ is a superoperator to describe an interaction between the working fluid and heat bath, and $k$ is either {\bf A} (Agarwal) or {\bf L} (Lindbladian).  Equation~\eqref{eq:general} without the superoperator is just a von Neumann equation, which describes a closed quantum system. Note that a hat symbol $(\hat{\cdot })$ denotes an operator.

The superoperator of the Agarwal bath~\cite{Agarwal2013quantum} is written as
\begin{align}
    {\mathcal{L}}_{_{\bf A}}(\hat\rho(t)) = 
     -\frac{i\kappa }{\hbar } [ \hat{x} , \{\hat p , \hat \rho(t)\}]
    -\frac{2\kappa m  \omega}{  \hbar }(\bar n + \frac{1}{2}) [\hat x,[\hat x ,\hat \rho(t)]].\label{eq:Agarwalbath}
\end{align}
where $\kappa$ is a heat conductance that governs the energy exchange rate between the working fluid and the Agarwal bath, and $\bar n$ is the expectation value of the number operator for the heat bath of temperature $T$, $\bar n = [\exp{(\hbar\omega/k_{B} T )}-1]^{-1}$.  Expanding Eq.~\eqref{eq:Agarwalbath} in the high-temperature limit, it becomes the Caldeira-Leggett master equation, which is well known to model quantum tunneling phenomena in a dissipative system~\cite{CALDEIRA1983374}.

For the Lindblad bath, it is as follows:
\begin{align}
   {\mathcal{L}}_{_{\bf L}}(\hat\rho(t)) = 
     \frac{\gamma}{2} ( \bar{n} + 1 )\left [\hat a \hat \rho (t) \hat a^{\dagger}- \frac{1}{2} [\hat a^{\dagger} \hat a    \hat\rho (t) + \hat \rho (t) \hat a^{\dagger}\hat a ]\right]\nonumber\\
    + \frac{\gamma}{2} \bar{n}  \left[\hat a^{\dagger}\hat \rho (t) \hat a - \frac{1}{2} [\hat a\hat a^{\dagger} \hat   \rho(t) +\hat \rho(t)\hat a\hat a^{\dagger} ]\right],
\label{eq:Lindbladbath}
\end{align}
where  $\gamma$ is heat conductance of the Lindblad bath~\cite{Breuer2006TheTheory} and $\hat{a}$ ($\hat{a}^{\dagger}$) represents an annihilation (creation) operator. The annihilation operator is the combination of position and momentum operators, $\hat a = \sqrt{\frac{m \omega}{2 \hbar}}(\hat x + \frac{i}{m\omega}\hat p )$, and the creation operator is the complex conjugate of $\hat a$, $\hat a^{\dagger} = \sqrt{\frac{m \omega}{2 \hbar}}(\hat x -\frac{i}{m\omega}\hat p)$.
For the adequate comparison of the Agarwal bath with the Lindblad bath, we set the heat conductance of the Agarwal bath as $\kappa=\gamma/8$.

In the adiabatic process, the volume of the working fluid is changed without heat transfer between heat bath and the working fluid, so that the master equation with $\gamma=0$ corresponds to the adiabatic process, where $\hat{H}(t)$ is explicitly time dependent and all the energy change of the working fluid becomes work. 

Combining these processes into a quantum Otto cycle, we generate the following procedure: 
First, we compress the working fluid in the adiabatic process, where work is exerted on the working fluid and its energy level becomes higher than it was before it was. Second, we connect the working fluid to a hot bath with temperature $T_{\rm h}$. In the hot isochore, heat is transferred to the working fluid from the hot bath, which is transformed as other types in the following adiabatic process. Third, in the adiabatic process, we expand the working fluid, so that the energy of the engine is transferred to the external agent. Finally, in the cold isochore, we connect the working fluid to the cold bath with temperature $T_{\rm c}$. 
Since the working fluid does not connect to the heat bath when Hamiltonian has the explicit time dependence, solving an Otto engine is easier than other finite-time cyclic heat engines. 

\subsection{Working fluid: Harmonic oscillator}
\label{working fluid}

To make our problem simple and analytically tractable, we employ harmonic oscillators as the working fluid of the Otto cycle. The harmonic oscillator is useful to model diverse phenomena, such as a cavity, a trapped ion, a RLC circuit, and a mechanical spring.
The Hamiltonian for the time-dependent harmonic oscillator is given by
\begin{align}
    \hat{H}(t) = \frac{ \hat{p}^2 }{2m} + \frac{m\omega^2(t)\hat{x}^2}{2},
    \label{eq:harmonic oscillator}
\end{align}
where $m$ and $\hat x $ ($\hat p$) are mass and position (momentum) operator, respectively. 
For the harmonic gas, it is known that the inverse of the frequency $\omega(t)$ corresponds to the volume of the working fluid~\cite{Romero-Roch2005Equation}. Hence, in the adiabatic process, we change the frequency  $\omega(t)$, whereas in the isochore, we do not.

With the Wigner function representation, we can map Eq.~\eqref{eq:general} for the density matrix to an equation for the $c$ number. The Wigner function describes a quasiprobability that represents the density function operator as a real function, which is written as
\begin{align}
    W(x,p) = \frac{1}{\pi\hbar}\int dz \, e^{ - 2 i p z /\hbar } \langle x + z | \hat{\rho}(t) | x-z \rangle.
\label{eq:Wignerdef}
\end{align}
The quasiprobability does not satisfy probability axioms and can have negative values. For the Gaussian state, $W(x,p)$ is guaranteed to be a non-negative value~\cite{gardiner2004quantum}.

For the harmonic oscillator,  the master equation of $W(x,p)$ is 
\begin{align}
    \partial_{t} W(x,p) = 
    -\vec\nabla_{q}
    \cdot
    [\mathcal{A}_{k}
    \cdot \vec{q}
    -
    \mathcal{B}_{k}
    \cdot
    \vec\nabla_{q}]
    W(x,p),
    \label{eq:Wignereq}
\end{align}
where
\begin{align}
    \mathcal{A}_{_{\bf A}} = \left(\begin{smallmatrix}
    0 &  \frac{1}{m}  \\
    -m \omega^2(t) & -\frac{\gamma}{4}
    \end{smallmatrix}\right); \quad
    \mathcal{B}_{_{\bf A}} =  \left(\begin{smallmatrix}
    0 & 0             \\
    0                & \frac{m \gamma  }{4} \tilde T
    \end{smallmatrix}\right)
\label{eq:AgarwalWignerMat}
\end{align} 
for the Agarwal bath, and
\begin{align}
    \mathcal{A}_{_{\bf L}} = \left(\begin{smallmatrix}
    -\frac{\gamma}{4} &  \frac{1}{m}  \\
    -m \omega^2(t) & -\frac{\gamma}{4}
    \end{smallmatrix}\right); \quad
    \mathcal{B}{_{\bf L}} =  \left(\begin{smallmatrix}
    \frac{\gamma \tilde T }{4 m \omega^2(t)} & 0             \\
    0                & \frac{m \gamma \tilde T }{4}
    \end{smallmatrix}\right)
\label{eq:LindWignerMat}\end{align} 
for the Lindblad bath, and  $\tilde T = \hbar \omega (\bar{n} + 1/2 )$.

Equation~\eqref{eq:Wignereq} has the same structure as the Fokker-Planck equation~\cite{Risken}. 
The corresponding Langevin equation to the master equation of the Wigner function is called the quasiclassical Langevin equation~\cite{gardiner2004quantum}.
The Langevin equation for the Agarwal bath is
\begin{align}
    \begin{split}
        \partial_{t} x &= \frac{p}{m} ,  \\
        \partial_{t} p &= -m\omega^2 x 
        - \frac{\gamma}{4} p + \sqrt{\frac{ \gamma \hbar m\omega (\bar{n} + 1/2 ) }{4}} \eta_{p}(t), 
    \end{split}
    \label{eq:langevinA}
\end{align}
where
   $\langle \eta_{i}(t) \eta_{j} (t') \rangle = 2\delta_{i,j}\delta(t-t')$.
In this case, if we take the high-temperature limit, then Eq.~\eqref{eq:langevinA} becomes the Langevin equations for a Brownian particle.

The Langevin equations for the Lindblad bath are 
\begin{align}
    \begin{split}
        \partial_{t} x &= \frac{p}{m} - \frac{\gamma}{4} x 
        + \sqrt{ \frac{\gamma \hbar (\bar{n} + 1/2 )  }{4 m\omega }  } \eta_{x}(t),   \\
        \partial_{t} p &= -m\omega^2 x 
        - \frac{\gamma}{4} p + \sqrt{\frac{ \gamma \hbar m\omega (\bar{n} + 1/2 ) }{4}} \eta_{p}(t). 
    \end{split}
    \label{eq:langevinL}
\end{align}
Note that, for the Lindblad bath, an additional heat channel exists in position. For the governing equation for momentum, both cases are exactly the same due to the choice of $\kappa = \gamma/8$, which helps to resolve the role of the positional heat channel in Eq.~\eqref{eq:langevinL}. Due to this fact, the relaxation of potential energy for the Lindblad bath and the Agarwal bath are quite different, which leads to huge difference in the performances of both Otto engines in finite time. Such outcomes are presented and discussed with possible origins in Sec.~\ref{result}.

Since Eq.~\eqref{eq:Wignereq} has the quadratic form, the cyclic steady state of Otto engines can be described by Gaussian. As a result, the Wigner function is non-negative in the limit cycle~\cite{Santos2017Wigner}. Due to the left-right symmetry for the breathing potential, $\langle \hat{x}\rangle$ and $\langle \hat{p} \rangle$ are zero in the cyclic steady state. Therefore, it is enough to calculate the second moments for describing cyclic steady states. 

With the adjoint master equation of $W(x,p)$, we can write down equations for Hamiltonian $\hat{H}$, Lagrangian $\hat{L}$, and a correlation function $\hat{D}$, respectively:
\begin{align}
\begin{split}
\hat H(t) &=\frac{\hat p^2}{2m} + \frac{m\omega^2(t) \hat{x}^2}{2},\nonumber\\
\hat{L}(t) &=\frac{\hat p^2}{2m} - \frac{m\omega^2(t) \hat{x}^2}{2},\nonumber\\
\hat{D}(t) & \equiv \frac{\omega(t) (\hat{x}\hat{p} + {p}\hat{x})}{2},\nonumber
\end{split}
\end{align}
which are the linear combinations of second moments. 

The evolution of a vector 
$$
\vec\phi(t)\equiv (\langle \hat{H}(t) \rangle, \langle \hat{L}(t) \rangle,\langle \hat{D}(t) \rangle,\langle \hat{I} \rangle )^{T}
$$ 
where $\hat I $ is the identity operator.
The vector is governed by a linear master equation~\cite{Rezek2006Irreversible}: 
\begin{align}
    \frac{\rm d}{{\rm d}t}
    \vec\phi(t)
    =
    \mathcal{M}^{k}_{j}
    \vec\phi(t),
\label{eq:adiabaticHLDI}
\end{align}
where $k$ is either {\bf A} or {\bf L}, and $j$ is either adiabatic ({\bf a}) or isochoric (${\bf i}\in \{\rm c, h\}$).

In the adiabatic process, the matrix $\mathcal{M}$ of Eq.~\eqref{eq:adiabaticHLDI} is written as
\begin{align}
    \mathcal{M}^{\bf A/L}_{\bf a} 
    =
    \omega(t)
    \begin{pmatrix}
    \frac{\dot \omega(t) }{\omega^2(t)} & - \frac{\dot{\omega}(t)}{\omega^2(t)} & 0 & 0  \\
    - \frac{\dot \omega(t) }{\omega^2(t)} &  \frac{\dot{\omega}(t)}{\omega^2(t)} & - 2 & 0  \\
    0 & 2 & \frac{\dot\omega(t)}{\omega^2(t)} & 0 \\
    0 & 0 & 0 & 0
    \end{pmatrix}.
\label{eq:adia_mat}
\end{align}
Here work per unit time is given as 
\begin{align}
    \partial_{t} \langle \hat{H}\rangle = \frac{\dot{\omega}(t)}{\omega(t)}(\langle \hat{H} \rangle - \langle \hat{L} \rangle )
    \label{eq:friction}.
    \end{align}    
On the right-hand side of Eq.~\eqref{eq:friction}, the second term associated with $\langle \hat{L}\rangle$ is called {\em friction} because it disappears    
in the quasistatic limit and decreases the power of Otto heat engines in the finite-time mode~\cite{Rezek2006Irreversible}. However, we show that in the engine with an Agarwal bath, the friction term can have the same sign as the first term, so that it helps to enhance the performance of the engine.

When $\dot\omega(t)/\omega^2(t)$ is constant, we can factor out $\omega(t)$ in the adiabatic matrix $\mathcal{M}_{\bf a}$ and the solution of Eq.~\eqref{eq:adiabaticHLDI} has a closed form~\cite{Insinga2018quantum}. 
\begin{align}
 \omega(t) = \frac{\omega_i \omega_f}{\omega_{f}-(\omega_{f}-\omega_{i})t/\tau},
\end{align}
where $i$ is initial, $f$ is final, and $\tau$ is the time of the adiabatic process. 
Then the propagator of the adiabatic process, $\mathcal{P}_{if}$, is written as 
\begin{align}
\ln{\left(\mathcal P_{if}\right)} = 
     \begin{pmatrix}
     r_{w} & -r_{w} & 0 & 0  \\
    - r_{w} & r_{w} & \frac{2\tau_{if} r_{w} }{\omega_{f}^{-1} - \omega_{i}^{-1} }  & 0  \\
    0 & -\frac{2\tau_{if} r_{w} }{\omega_{f}^{-1} - \omega_{i}^{-1} }  & r_{w} & 0 \\
    0 & 0 & 0 & 0
    \end{pmatrix},
\end{align}
where $ r_{w} \equiv \ln \left({\omega_f/\omega_i}\right)$.
For the simplicity, we take the notation of the propagator of the adiabatic compression (expansion) process as $\mathcal P_{\rm ch}$ ($\mathcal P_{\rm hc}$) as stated in Fig.~\ref{fig:conceptual}. 

In the isochore, the matrix $\mathcal{M}$ of Eq.~\eqref{eq:adiabaticHLDI} is given as \begin{align}
    \mathcal{M}^{\bf A}_{\bf i}
    =
    \begin{pmatrix}
    -\frac{\gamma }{4} & -\frac{\gamma }{4} & 0 & \frac{\gamma  \tilde T_{\bf i} }{4}\\
    -\frac{\gamma }{4} &  -\frac{\gamma }{4} & - 2\omega_{\bf i} & \frac{\gamma \tilde T_{\bf i} }{4}  \\
    0 & 2\omega_{\bf i} & -\frac{\gamma }{4} & 0 \\
    0 & 0 & 0 & 0
    \end{pmatrix}
\label{eq:ciso_mat}
\end{align}
and
\begin{align}
    \mathcal{M}^{\bf L}_{\bf i}
    =
    \begin{pmatrix}
    -\frac{\gamma }{2} & 0  & 0 & \frac{\gamma \tilde T_{\bf i} }{2}  \\
    0 &  -\frac{\gamma }{2} & - 2\omega_{\bf i} &  0  \\
    0 & 2\omega_{\bf i}          & -\frac{\gamma }{2} & 0 \\
    0 & 0 & 0 & 0
    \end{pmatrix}
\label{eq:qiso_mat}
\end{align}
where {\bf i} is {h} ({c}) for the hot (cold) isochore.
Because the matrix in the isochore is independent of time, the propagator is given as $\mathcal{P}^{k}_{\bf i } = \exp{(\mathcal{M}^{k}_{\bf i }t )} $. 
By substituting the matrix in Eq.~\eqref{eq:adiabaticHLDI} with Eqs.~\eqref{eq:ciso_mat} and \eqref{eq:qiso_mat}, we get equations for the evolution of the Hamiltonian, Lagrangian and correlation. For the case of $\mathcal{M}^{\bf L}_{\bf i}$, $\langle \hat{H}\rangle$ directly approaches to energy in equilibrium and does not couple to $\langle \hat{L}\rangle$ and $\langle \hat{D}\rangle$. On the other hand, for the case of $\mathcal{M}^{\bf A}_{\bf i}$, 
they are coupled to one another. This difference leads to a big difference in the performance of finite-time engines in cyclic steady states, which is discussedin Sec.~\ref{result} in detail. 

Rearranging equations for the isochores, we obtain the equations for kinetic energy (KE) and potential energy (PE). The dynamic equations of KE and PE are written as 
\begin{align}
  \begin{split}
    \frac{\rm d}{{\rm d}t} \langle \frac{\hat p ^2 }{2m}\rangle &= - \frac{\gamma}{2} \langle   \frac{\hat p ^2 }{2m}\rangle - \omega_{\bf i} D + \frac{\gamma \tilde T_{\bf i}}{4}, \\
    \frac{\rm d}{{\rm d}t} \langle \frac{m \omega^2_{\bf i} \hat x ^2 }{2}\rangle &=  \omega_{i} D.
  \end{split}   
  \label{eq:PEKE_A}
 \end{align}
for the Agarwal bath and
\begin{align}
  \begin{split}
    \frac{\rm d}{{\rm d}t} \langle \frac{\hat p ^2 }{2m}\rangle &= - \frac{\gamma}{2} \langle \frac{\hat p ^2 }{2m}\rangle - \omega_{\bf i} D + \frac{\gamma \tilde T_{\bf i}}{4},\\
    \frac{\rm d}{{\rm d}t} \langle \frac{m \omega^2_{\bf i} \hat x ^2 }{2}\rangle &=  \omega_{\bf i} D - \frac{\gamma}{2} \langle \frac{m \omega^2_{\ bf i} \hat x ^2 }{2}\rangle  + \frac{\gamma \tilde T_{\bf i}}{4}.
 \end{split}
 \label{eq:PEKE_Lind}
\end{align}
for the Lindblad bath.
For both cases, the governing equation for KE is the same and this is our criterion to regulate a heat conductance for both baths.
%
From propagator expressions as shown in Fig.~\ref{fig:conceptual}, we are able to calculate cyclic steady states and the performance of engines regarding the assigned bath. The propagator for one cycle is given by $\mathcal{P}^{k}_{\rm cyc} \equiv \mathcal P^{k}_{\rm c} \mathcal{P}_{\rm hc} \mathcal P^{k}_{\rm h} \mathcal{P}_{\rm ch}$. 

With the condition that the Hamiltonian, Lagrangian and the correlation function remain the same after one cycle, the cyclic steady state $\vec \phi_{\rm ss}^{k}$ can be calculated~\cite{Insinga2018quantum}. 
Then, work and heat are written as 
\begin{align}
    \begin{split}
        \mathcal{W}_{\rm ch}^{k} &= \vec d \cdot ( \mathcal P_{\rm ch} - \mathcal I ) \cdot \vec \phi_{\rm ss}^{k} \\
        \mathcal{W}_{\rm hc}^{k} &= \vec d \cdot (\mathcal P_{\rm hc } - \mathcal{I} ) \mathcal P_{\rm h}^{k} \mathcal P_{\rm ch}  \cdot \vec \phi_{\rm ss}^{k} \\
        Q_{\rm h}^{k} &= \vec d \cdot ( \mathcal P^{k}_{\rm h} - \mathcal I ) \mathcal P_{\rm ch} \cdot \vec \phi_{\rm ss}^{k} \\
        Q_{\rm c}^{k} &= \vec d \cdot ( \mathcal P_{\rm c}^{k} - \mathcal I ) \mathcal P_{\rm hc} \mathcal P^{k}_{\rm h} \mathcal P_{\rm ch} \cdot \vec \phi_{\rm ss}^{k},
    \end{split}
    \label{eq:workandheat}
\end{align}
where 
\begin{align}
    \vec d \equiv (1,0,0,0)^T
\end{align}
and $\mathcal I$ is an identity matrix of size four. 
From Eq.~\eqref{eq:workandheat}, the performance of the Otto engine, its efficiency and power, can be calculated as follows:
\begin{align}
    \begin{split}
        \eta^{_{k}}&= - (\mathcal{W}^{k}_{\rm ch} + \mathcal{W}^{k}_{\rm hc} )/Q_{\rm h}^{k},\\ 
        P^{_{k}}&= - (\mathcal{W}^{k}_{\rm ch} + \mathcal{W}^{k}_{\rm hc} )/\tau_{\rm cyc}.
    \end{split}
    \label{eq:effiandpower}
\end{align}

\section{Results}
\label{result}

In this section, we present the remarkable difference between Agarwal and Lindbladian Otto engines, in the context of the performance of the finite-time engine, which is based on exact solutions. However, the exact mathematical forms are not directly shown in this paper since they are quite complicated.  Instead, we present the analytic forms of the approximated result in the short cycle-time limit. Using the analytic condition for the divergence of the engine with the resonance, we show that the finite-time Otto heat engine is different from the quasistatic limiting case. For the finite-time performance, we provide enumerated results to support our interesting findings, where we set all the parameters to be dimensionless, and for the simplicity, $k_{B}=\hbar=1$.

Before moving onto our results, we briefly review the quasistatic behavior and give some intuition of the Otto heat engine. When the time periods of both adiabatic processes are sufficiently large, the engine has the universal efficiency, $\eta_{_{\rm O}} = 1- \omega_{\rm c}/\omega_{\rm h}$, known as the quantum Otto efficiency~\cite{Rezek2006Irreversible}. The Otto efficiency is smaller than the Carnot efficiency $\eta_{_{\rm C}}=1-T_{\rm c}/T_{\rm h}$.  This statement is consistent with the fact that the system operates as an engine only when $T_{\rm c}/T_{\rm h} < \omega_{\rm c}/\omega_{\rm h}$. In the quasistatic limit, if we control the frequency ratio beyond it, the Otto cycle becomes a refrigerator, rather than a heat engine~\cite{Kosloff2017-QuantumOtto}.

\begin{figure}
 \includegraphics*[width=\columnwidth]{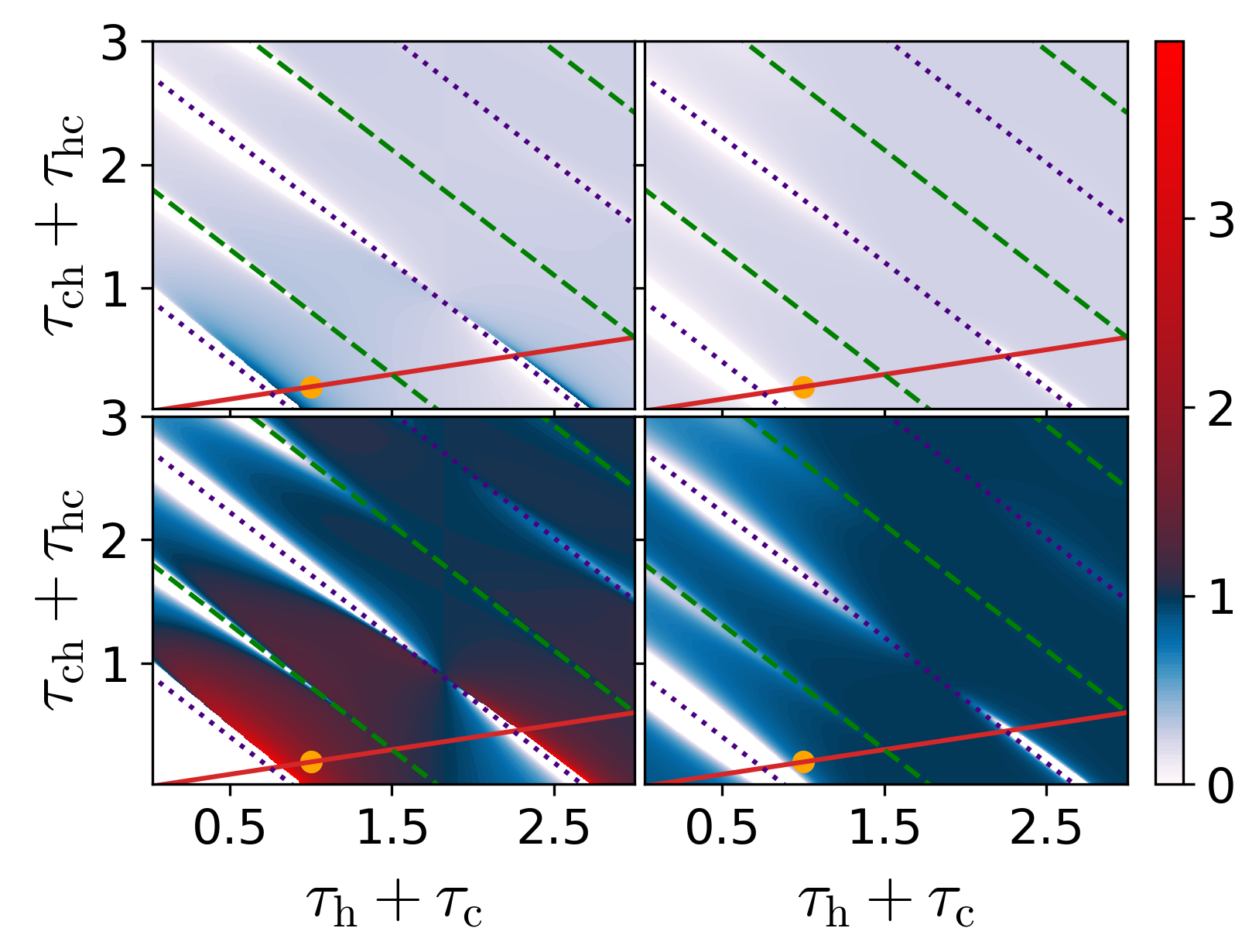}
\caption{ The contour plots of $\eta^{_{\bf A}}$(upper left), $\eta^{_{\bf L}}$(upper right), $\eta^{_{\bf A}}/\eta_{\rm O}$(lower left) and $\eta^{_{\bf L}}/\eta_{\rm O}$(lower right) are presented as functions of $\tau_{\rm h}+\tau_{\rm c}$ (the sum of the isochoric time, $x$ axis) and $\tau_{\rm ch} +\tau_{\rm hc}$ (the sum of the adiabatic time, $y$ axis). 
Since we plot only when the cycle behaves as a heat engine, there are diagonal blank spaces.
Unless isochoric process time is long, the blank spaces well coincide with purple dotted (green dashed) lines which are derived from Eq.~\eqref{eq:resonancecond} when $n$ is odd (even).
Near purple resonance lines, we can find some regions that show $\eta^{_{\bf A}}\gg\eta^{_{\bf L}}$. In both panels, the red solid line represents $(\tau_{\rm ch}+\tau_{\rm hc})/(\tau_{\rm h}+\tau_{\rm c})=1/5$, and the orange dot corresponds to the case of $\tau_{\rm cyc}=1.2$, which is discussed in Figs.~\ref{fig:efficiencyandpower} and \ref{fig:trajectory}. Here we set all parameters to be dimensionless and $\hbar= k_{B} = m=\gamma = 1$, $\omega_{\rm h}=4, \omega_{\rm c}=3, T_{\rm h} = 200$,  and $T_{\rm c} = 1$, which are kept used from now on unless other values are indicated explicitly. For simplicity, we choose $\tau_{\rm h}=\tau_{\rm c}$ and $\tau_{\rm ch}=\tau_{\rm hc}$.}
\label{fig:time2dimcontour}
\end{figure}

In Fig.~\ref{fig:time2dimcontour}, we show how the Agarwal (Lindbladian) Otto engine in the left (right) panel works with the following parameter settings: $m=\gamma = 1,  \omega_{\rm h}=4, \omega_{\rm c}=3, T_{\rm h} = 200$, and $T_{\rm c} = 1$. The x-axis (y-axis) is the sum of two isochoric (adiabatic) times, and we plot the efficiency only when the engine behaves as a heat engine. 
Blank spaces appear along the dotted lines, which are drawn by the divergence or resonance condition by Eq.~\eqref{eq:resonance}. The condition is based on the classical argument, {\it if the period of the system is a multiple of the period of the driving force, then resonance can be observed}. Due to the left-right symmetry of our engines, we can observe resonances even when the half period of the system is a multiple of the period of the driving force. With this condition and the lack of dissipation to the heat bath, the energy of the working fluid can be accumulated in every cycle, which leads to the energy divergence. Hence, in this case, the cyclic steady state does not exist. For the quantum Otto heat engine, the resonance condition neglecting the effect of the heat bath is calculated as follows: 
\begin{align}
        n \pi =& \int^{\tau_{\rm cyc }}_{0} {\rm d}t \, \omega(t) 
\label{eq:resonance} 
\end{align}
The simplified resonance condition can be written as
\begin{align}
        n \pi = \omega_{\rm c}\tau_{\rm c} + \omega_{\rm h}\tau_{\rm h} 
        + \frac{\omega_{\rm c}\omega_{\rm h}}{\omega_{\rm h} - \omega_{\rm c}}\ln{\left( \omega_{\rm h}/\omega_{\rm c} \right)}(\tau_{\rm ch} + \tau_{\rm hc} ).
\label{eq:resonancecond} 
\end{align}
The right-hand side of Eq.~\eqref{eq:resonancecond} is the summation of phase difference for the four processes in the Otto cycle. Near the condition of Eq.~\eqref{eq:resonancecond} in the short-time region, the working fluid continuously gets energy, so that energy diverges. However, if the contact time with the heat bath is long enough to be dissipated, then energy does not pile up in the working fluid and a cyclic steady state exists. 

\begin{figure}
 \includegraphics*[width=0.8\columnwidth]{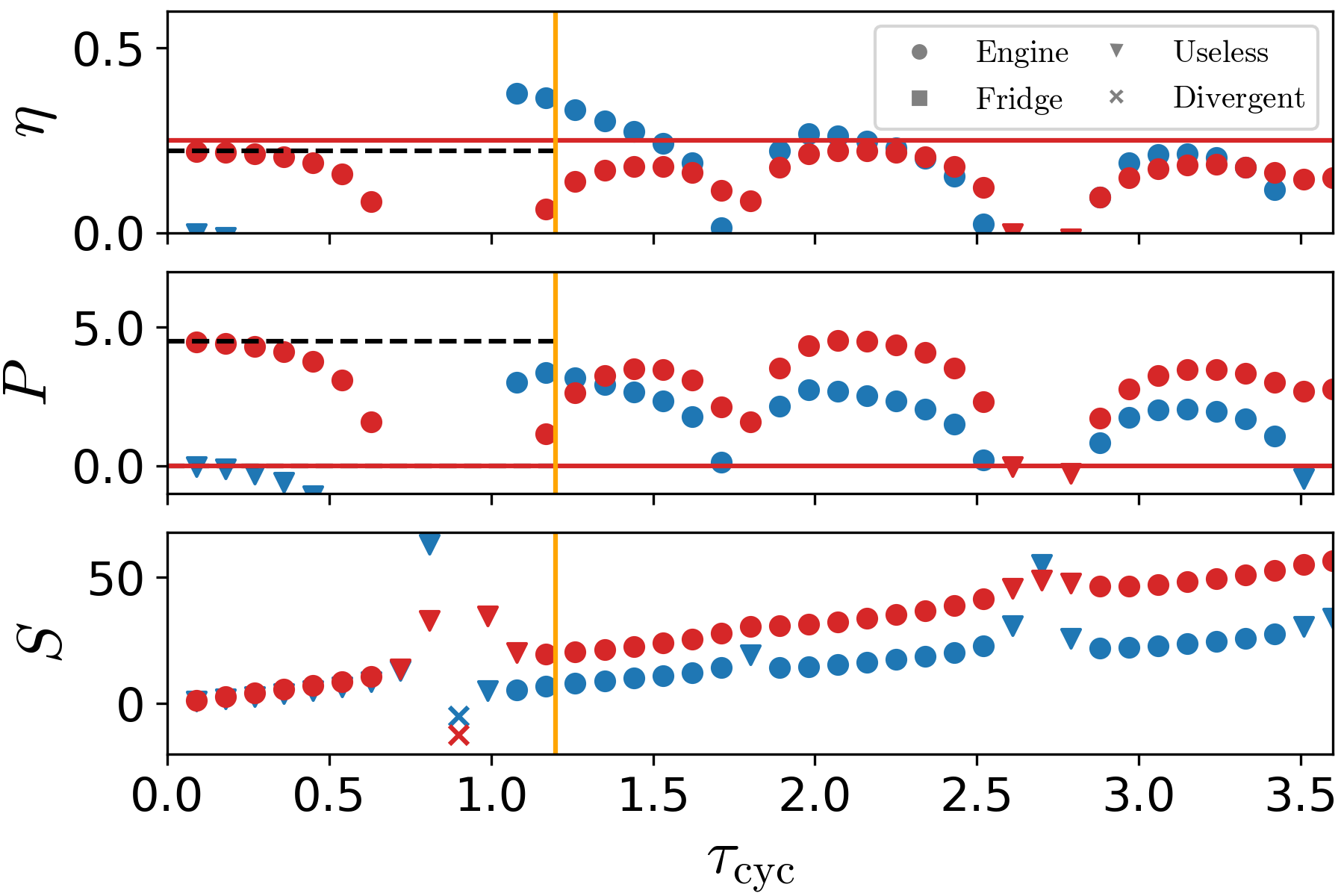}
\caption{Along the red line of each panel in Fig.~\ref{fig:time2dimcontour}, we compare the performance of the Agarwal Otto engine (blue, {\color{blue} $\bullet$}) with that of the Lindbladian (red, {\color{red} $\bullet$}), in the context of its efficiency $\eta$ (top), power $P$ (middle), and entropy $S$ (bottom), which are as a function of $\tau_{\rm cyc}$. In the quasistatic limit ($\tau_{\rm cyc}\to \infty$), $\eta^{_{\bf A/L}}\to \eta_{_ {\rm O}}$ and $P^{_{\bf A/L}}\to 0$, which are drawn as horizontal red solid lines. In addition, the analytic short-time results of Eq.~\eqref{eq:shorttimework} for $\eta^{_{\bf L}}$ and $P^{_{\bf L}}$ are drawn as horizontal black dashed lines up to $\tau_{\rm cyc}=1.2$, whereas $\eta^{_{\bf A}}\to 0$ and $P^{_{\bf A}}\to 0$. Vertical orange solid lines are drawn at $\tau_{\rm cyc}=1.2$ (orange dots in Fig.~\ref{fig:time2dimcontour}), where $\eta^{_{\bf A}} > \eta_{\rm O} >\eta^{_{\bf L}}$. For nonengine or unphyiscal values, we use different symbols from that of the heat engine and put some explanations as keys: fridge, useless, and divergent.}
\label{fig:efficiencyandpower}
\end{figure}

In Fig.~\ref{fig:efficiencyandpower}, we show the performance of two finite-time Otto engines along the red line of each panel in Fig.~\ref{fig:time2dimcontour}, where the ratio of an isochoric time to an adiabatic time is fixed as $5:1$. 
The finite-time quantum Otto cycle can be one of the following four ways:
 In the heat engine, heat flow is converted to work.  In the refrigerator, heat is absorbed from the cold bath due to work. In the useless machines, both work and heat are consumed and exerted into the cold bath. 
We allocate different symbols to each case, circles for engines, squares for refrigerators, triangles for useless machines, and crosses for the divergent case in Fig.~\ref{fig:efficiencyandpower}. However, the refrigerator is not found with those parameters. We also present the behavior of entropy for both cases, which shows that the entropy in the short-time limit is the same, but the Lindbldian case is larger than the Agarwal case in the finite-time mode. The quasistatic values of the efficiency and the power are plotted as a red horizontal line in Fig.~\ref{fig:efficiencyandpower}. In the Appendices,  we confirm that both Otto engines approach the quasistatic values with oscillatory behavior (see Fig. A1 in Appendix~\ref{appendix-A}).

A noticeable difference between the Agarwal Otto engine and the Lindbladian Otto engine is that the efficiency of the Agarwal case is higher than that of the Lindbladian case near the resonant condition from Eq.~\eqref{eq:resonancecond}. 
To figure out the origin of such a notable difference, we measure trajectories of the KE, PE, Hamiltonian and the friction term of the limit cycle at $\tau_{\rm cyc}=1.2$ when the difference is dominant. 
It is because they are essential to calculate the performance of two heat engines.

\begin{figure}
 \includegraphics*[width=0.8\columnwidth]{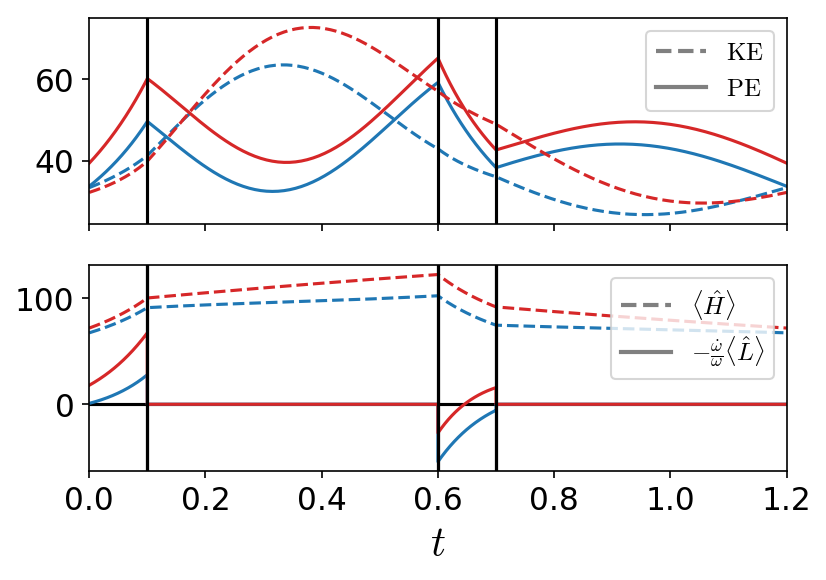}
\caption{ At $\tau_{\rm cyc}=1.2$ when $\eta_{_{\rm A}} > \eta_{_{\rm O}}$ (indicated in Fig.~\ref{fig:efficiencyandpower}),  the expectation values of the kinetic energy (KE, dashed lines) and the potential energy (PE, solid lines) are plotted as a function of time $t$, where we set $\tau_{\rm h}=0.1$ and $\tau_{\rm hc}=0.5$. Three vertical black solid lines represent three boundaries, from the adiabatic compression to the hot isochore, from the hot isochore to the adiabatic expansion, and  from the adiabatic expansion to the cold isochore, respectively (from the left to the right). Here we use the same parameters and colors as those used in Fig.~\ref{fig:efficiencyandpower}. In  Agarwal's adiabatic expansion (the last part for $0.7\le t\le 1.2$, see the two blue lines), the PE is always larger than the KE, which is different from the Lindbladian where the sign of the Lagrangian changes. This implies that the friction term, $-\frac{\dot\omega(t)}{\omega(t)}\langle \hat L \rangle$ in Eq.~\eqref{eq:friction}, contributes to $\eta^{_{\bf A}} > \eta_{_{\rm O}}$.}
\label{fig:trajectory}
\end{figure}

\begin{figure}[t]
 \includegraphics*[width=0.8\columnwidth]{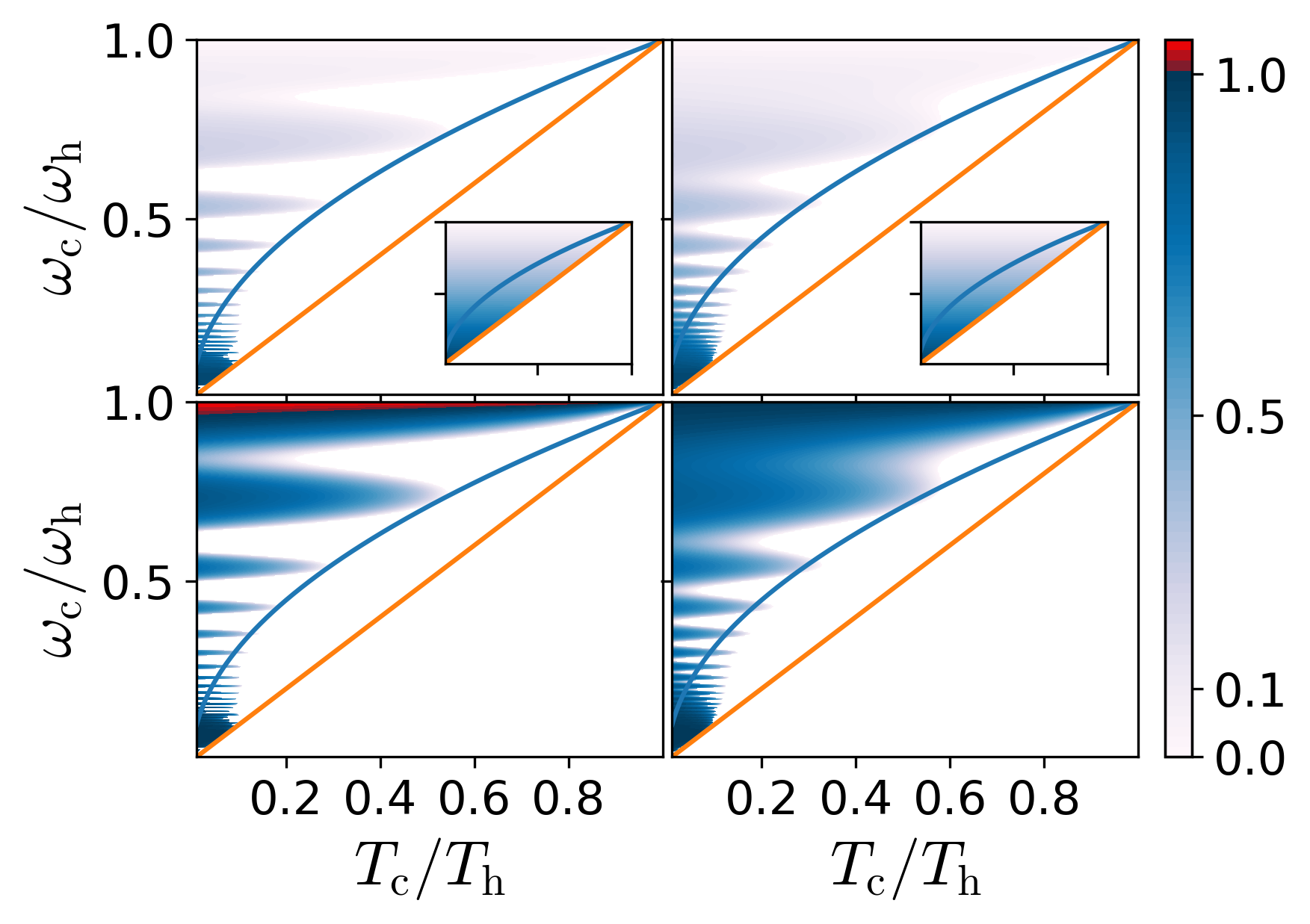}
\caption{The contour plots of $\eta^{_{\bf A}}$ (upper left), $\eta^{_{\bf L}}$ (upper right), $\eta^{_{\bf A}}/\eta_{\rm O}$ (lower left) and $\eta^{_{\bf L}}/\eta_{\rm O}$ (lower right) are shown as a function of $T_{\rm c}/T_{\rm h}$ (temperature ratio, x-axis) and $\omega_{\rm c}/\omega_{\rm h}$ (frequency ratio, y-axis). Here most parameters are the same as before, but we change $\omega_{\rm c} = 3, T_{\rm c} = 100, \tau_{\rm c}=\tau_{\rm h} = 2$ and $\tau_{\rm ch}=\tau_{\rm hc} = 0.4$, which nicely show how our enumeration result are bounded by the condition derived in the high-temperature limit. 
Blue (orange) guided lines are the boundaries between the heat engine and the others such as the refrigerator and the useless machine in the short-time (quasistatic) limit. 
The short-time limit was obtained from the Lindbladian work expression of Eq.~\eqref{eq:shorttimework} in the high-temperature limit.  
When the frequency of the harmonic oscillator gets higher, the approximation of the short-time limit cannot be valid anymore. Therefore, $\eta^{_{\rm A}}$ and $\eta^{_{\rm L}}$ near the small frequency ratio can be inbetween the blue lines and the orange lines. The insets correspond to the quasistatic limit, where both cases show the same results.}
\label{fig:frequencytemp}
\end{figure}

In the upper panel of Fig.~\ref{fig:trajectory}, we present the expectation values of KE and PE for each engine at $\tau_{\rm cyc}=1.2$ (orange vertical line in Fig.~\ref{fig:efficiencyandpower}), where the solid (dashed) line is the PE (KE). In the lower panel of Fig.~\ref{fig:trajectory}, we present the expectation values of Hamiltonian and the frictional term for two engines.
It is observed that for the Agarwal Otto engine (red), the PE is always larger than the KE in the adiabatic expansion process ($0.6 \leq t \leq 0.7$). This means that the sign of the friction term ($-\frac{\dot\omega}{\omega}\langle \hat L\rangle$) is negative during the adiabatic expansion process and increases the extracted energy. 
As a result, the friction term increases the efficiency of the finite-time Agarwal Otto engine to exceed $\eta_{_{\rm O}}$. For the case of the Lindblad Otto engine, we have to invest more energy during the adiabatic compression process ($0.0\leq t \leq 0.1$) because the friction of the Lindblad engine is higher than of the Agarwal Otto engine. Note that in the quasistatic limit, the friction term becomes zero (see Fig. A2 in Appendix~\ref{appendix-B}, which shows additional trajectories for the three choices of $ \tau_{\rm cyc}$ from short-time to long-time regimes). The contrast of imbalance between KE and PE in expansion process originates from the different dynamics for PE the in the isochore, Eqs.~\eqref{eq:PEKE_A} and \eqref{eq:PEKE_Lind}. Owing to the different relaxation behavior of the PE, the short-time performances of the heat engines also show immense differences. 

Another interesting phenomenon is observed for the very short cycle time, $\tau_{\rm cyc}\ll1$, where the Lindbladian Otto cycle can work as a heat engine but the Agarwal one cannot. 
For small $\tau_{\rm cyc}$, we approximate work in Eq.~\eqref{eq:workandheat} under the condition when the adiabatic time is shorter than the isochoric time and the expansion time equals the compression time for the simple result ($\tau_{\rm h},\tau_{\rm c}\gg\tau_{\rm ch} = \tau_{\rm hc}$).
 The first-order expressions of work are as follows: 
\begin{align}
    \mathcal{W}^{_{\bf A}} &= 0 + \mathcal{O}(\tau_{\rm cyc}^3)
    , \nonumber\\
    \mathcal{W}^{_{\bf L}} &=  
    \frac{\gamma  \tau_{\rm c} \tau_{\rm h} \left(\omega_{\rm h}^2-\omega_{\rm c}^2\right) \left(\tilde T_{\rm h} \omega_{\rm c}^2- \tilde T_{\rm c} \omega_{\rm h }^2\right)}{4 \omega_{\rm c}^2 \omega_{\rm h }^2 (\tau_{\rm c} + \tau_{\rm h })} + \mathcal{O}(\tau_{\rm cyc}^2).
    \label{eq:shorttimework}
\end{align}
When $\tau_{\rm h} = \tau_{\rm c}$, Lindblad work $\mathcal{W}^{_{\bf L} }$ in the complete sudden limit was calculated in the review paper by Kosloff and Rezek~\cite{Kosloff2017-QuantumOtto}. 
Using Eq.~\eqref{eq:shorttimework}, the efficiency and power values of two engines are calculated as well. 
For the Agarwal Otto engine, it is found that 
$\eta^{\bf A}$, $P^{\bf A} \rightarrow 0$  because $Q_{\rm h}^{\bf A}=\frac{\gamma\tau_{\rm c}\tau_{\rm h} (\tilde T_{\rm c} - \tilde T_{\rm h}) }{4 (\tau_{\rm c} + \tau_{\rm h} )}$ and the first-order term of $\mathcal W^{\bf A}$ is zero. So the Agarwal Otto cycle cannot be a heat engine in the short-time limit, which is true even when $\tau_{\rm ch}\neq \tau_{\rm hc}$. 
For the Lindbladian Otto engine, $\mathcal{W}^{_{\bf L}}$ is linear in $\tau_{\rm cyc}$, so that $P^{_{\bf L}}$, $\eta^{_{\bf L}}$ are nonzero, finite, and positive when $\omega_{\rm c }/\omega_{\rm h}> (\tilde T_{\rm c}/ \tilde T_{\rm h })^{1/2}$, which are shown in Fig.~\ref{fig:efficiencyandpower} as black dashed lines for the Lindblad case~\cite{Kosloff2017-QuantumOtto} [see Eqs. (A1) and (A2) in Appendix~\ref{appendix-C}  for the detailed mathematical expressions of $\eta^{_{\bf L}}$ and $ Q_{\rm h}^{\bf L}$].

In the high-temperature (classical) limit, such a condition becomes $\omega_{\rm c}/\omega_{\rm h}>(T_{\rm c}/T_{\rm h})^{1/2}$ as plotted in Fig.~\ref{fig:frequencytemp} with blue curved lines. So the valid parameter region in finite-time Otto cycles gets smaller than that in the quasistatic limit case ($\omega_{\rm c}/\omega_{\rm h} > T_{\rm c}/T_{\rm h}$), which is drawn by orange diagonal lines to guide the eye in Fig.~\ref{fig:frequencytemp}. 

Figure~\ref{fig:frequencytemp} shows wavy patterns because of resonance phenomena, which are the same as Fig.~\ref{fig:time2dimcontour}, and insets correspond to $\eta^{k}$ in the quasistatic limit. Note that blue lines  are derived from the Lindbladian case, but they fit quite well to the Agarwal case, too. This implies that the adiabatic process strongly relates to the boundary condition rather than the isochore. When the frequency of the working fluid is high, the short-time approximation ($\tau < \omega^{-1}, \gamma^{-1} $) fails, so that data points can exist over blue lines. In the region where $T_{\rm c}/T_{\rm h}>\omega_{\rm c}/\omega_{\rm h}$, both engines can work as a refrigerator with similar wavy patterns of cooling coefficient (see Fig.~A3 in Appendix~\ref{appendix-D}).  

\section{conclusion} 
\label{conclusion}

We have investigated the role of friction in quantum Otto engines with two different types of equilibrium heat baths, namely the Agarwal Otto engine and the Lindblad Otto engine. In the isochore, two master equations governing the dynamics are different. With the adjoint master equation for the Wigner function, up to  the second moments, they were exactly derived to solve the performances of the engines with a specific protocol. 

Based on our derivation of resonance conditions for both engines, it is found that the Agarwal Otto engine can exceed the quasistatic Otto efficiency in the finite-time mode. This is remarkably different from the Lindblad Otto engine near resonance conditions, which is also counterintuitive because there is positive feedback caused by friction.  Moreover, in the short cycle-time limit ($\tau_{\rm cyc} \rightarrow 0$), we were also able to derive the approximated expressions of work, which show that the Lindbladian can have nonzero power, which differs from the Agarwal Otto engine. It is because the Lindblad bath can directly transfer energy to the potential energy, so that the Otto cycle can directly extract energy from the potential energy in the short-time limit. 

Finally, in the finite-time mode, the power of the Lindblad engine is higher than that of the Agarwal engine, and its nondivergent parameter region is larger than that of the Agarwal engine. Such differences originate from the existence of the positional heat channel, which alters the relaxation behavior of the potential energy. Possible realizations of our work and implications of the frictional effect remain interesting subjects for future studies.

\begin{acknowledgments} 
This study was supported by
Basic Science Research Program through the National
Research Foundation of Korea (NRF) (KR) [NRF-2017R1D1A3A03000578 (S.L., M.H.); NRF-2017R1A2B3006930 (S.L., H.J.)], 
and a KIAS Individual Grant No. PG074001 (J.-M.P.) at Korea Institute for Advanced Study. 
We thank Ronnie Kosloff, Jinfu Chen, Yu-Han Ma, and Hyun-Myung Chun for fruitful comments and references. M.H. would also like to acknowledge the kind hospitality of Marcel den Nijs in the Physics Department at the University of Washington, Seattle, where this work was finalized for the sabbatical year.
\end{acknowledgments}

\onecolumngrid
\appendix

\setcounter{figure}{0}
\setcounter{table}{0}
\makeatletter
\renewcommand{\theequation}{A\arabic{equation}}
\renewcommand{\thefigure}{A\arabic{figure}}

\section{Long-Time Behavior} 
\label{appendix-A}

\begin{figure}[hb]
 \includegraphics*[width=0.5\columnwidth]{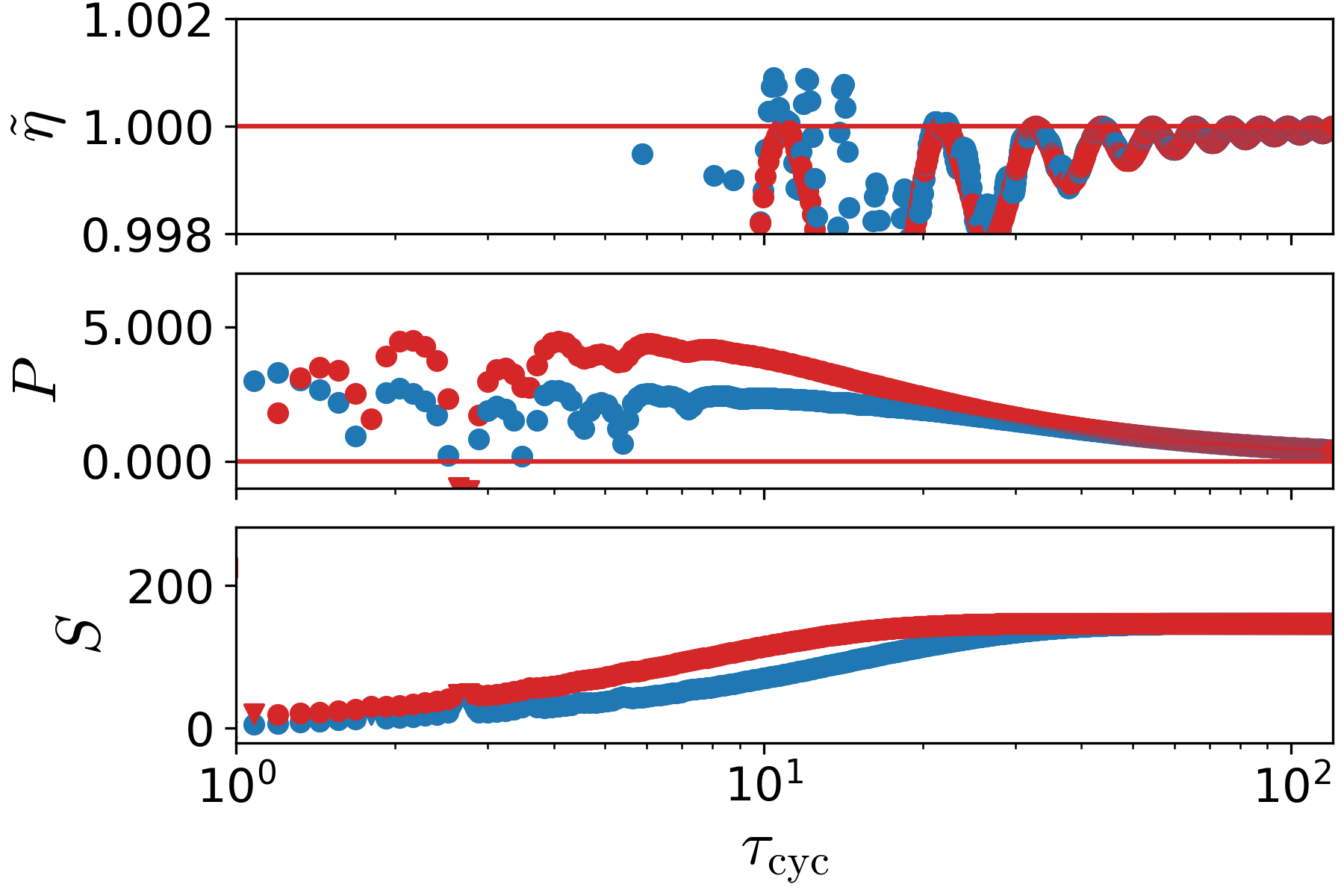}
\caption{Numerical data are taken with same parameters, same symbols,  and same colors  of Fig. 3 in the main text.}
\label{fig:long}
\end{figure}

To verify the long-time behaviors of two quantum Otto engines, we provide Fig.~\ref{fig:long} 
that shows the normalized efficiency $\tilde{\eta}\equiv \eta /\eta_{_{\rm Otto}}$, 
the power $P$, and the entropy production $S$.  As expected, $\tilde{\eta}\to 1 
\mbox{and}\ P\to 0$ for both cases. 

It is also found that two engines exhibit oscillatory behaviors to approach the limiting values.  For the case of the Lindbladian, such oscillatory behaviors near the quasi-static limit has been discussed in the recent study~\cite{Chen2019Boosting}. Because of the irreversibility of the isochore, the entropy production $S$ converges to some non-zero value. 

\section{Trajectories of Three Different Cycle Times:\\ From Short-time to Long-time}
\label{appendix-B}

\begin{figure}[hb]
 \includegraphics*[width=0.325\columnwidth]{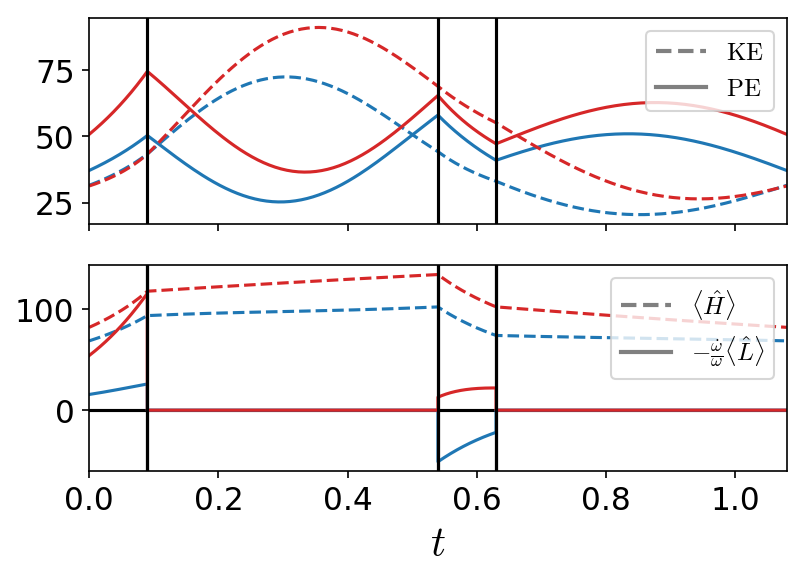}
 \includegraphics*[width=0.325\columnwidth]{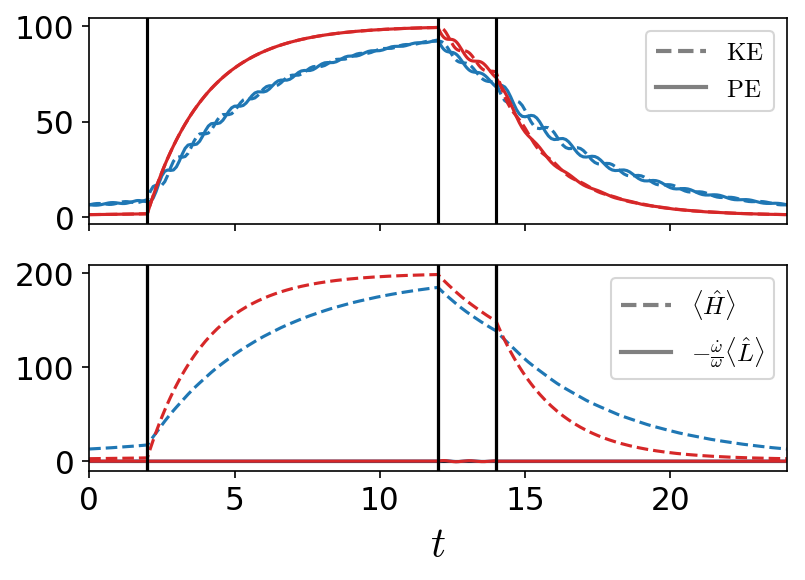}
 \includegraphics*[width=0.325\columnwidth]{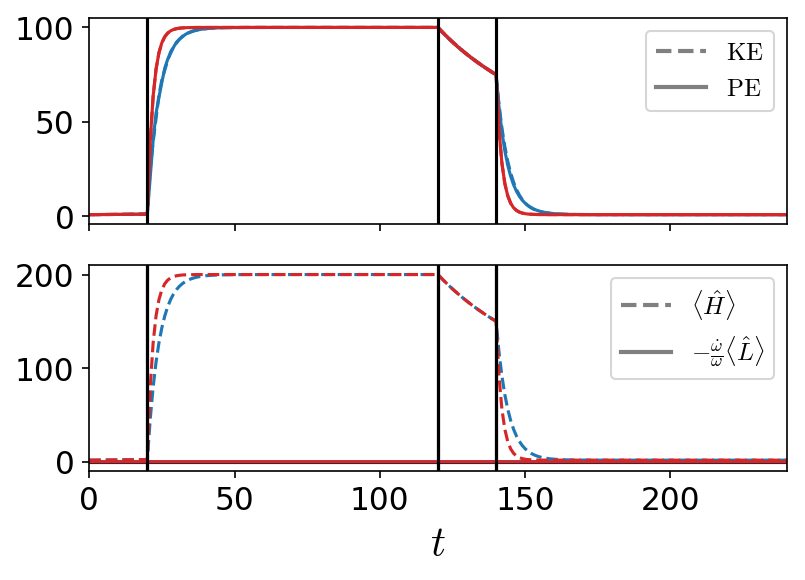}
 \caption{Trajectories of two quantum Otto engines with the Lindblad bath (red lines) and the Agarwal bath (blue lines) are presented at $\tau_{\rm cyc} = 1.08$ (left),  $\tau_{\rm cyc}=24$ (middle), and $\tau_{\rm cyc}=240$ (right), respectively. In the short-time limit, the difference of the frictional contribution becomes dominant. For the case of the Agarwal bath at $\tau_{\rm cyc}=1.08$, the contribution of $-\frac{\dot{\omega}}{\omega}\langle L\rangle$ is negative, which implies that the possibility of $\eta^{_{\bf A} } >\eta_{_{\rm Otto}}\ge \eta^{_{\bf L}}$ in the short-time limit as discussed in Fig. 4 of the main text.}
\label{fig:trajectories}
\end{figure}

We present trajectories of three different cycle times in Fig.~\ref{fig:trajectories}, where we plot the expectation values of kinetic energy (KE), potential energy (PE), Hamiltonian $\langle \hat{H}\rangle$, and the friction term $-\frac{\dot{\omega}}{\omega}\langle \hat{L}\rangle$ with the same parameters of Fig. 4 of the main text ($\hbar= k_{B} = m=\gamma = 1$, $\omega_{\rm h}=4, \omega_{\rm c}=3, T_{\rm h} = 200$,  and $T_{\rm c} = 1$).
For the case of $\tau_{\rm cyc}=240(\gg1)$ in the rightmost panel of Fig.~\ref{fig:trajectories}, it is found that for both engines initial and final value of each cycle are the same but the relaxation speed is different and the frictional effect completely disappears. As $\tau_{\rm cyc}$ decreases, the difference of two engines and the frictional effect appear. In the short-time limit, they become much clearer. In the long-time limit, two engines show the same performance as expected. However,  it seems that the additional heat channel in the position component of Lindblad bath (see Eq. (10) of the main text, compared to Eq. (9) of the main text) results in the faster relaxation of the Lindbladian (red lines in Fig.~\ref{fig:trajectories}) than that of the Agarwal bath (blue lines in Fig.~\ref{fig:trajectories}).  

\section{Short-Time Behavior}
\label{appendix-C}

For the case of Lindblad Otto engine, we can calculate the short-time behavior of the efficiency $\eta^{_{\bf L}}$, which is drawn in Fig. 3 of the main text as a horizontal black dashed line. When $1 \gg t = \tau_{\rm h} (= \tau_{\rm c }) \gg \tau_{\rm ch }(= \tau_{\rm hc})$, the efficiency $\eta^{_{\bf L}}$ and hot heat $Q_{\rm h}$ are written as follows:   

\begin{equation}
    \eta^{_{\bf L}} = 
    \frac{\left(\omega_{\rm c}^2-\omega_{\rm h}^2\right) \left[\gamma^2+8 \left(\omega_{\rm c}^2+\omega_{\rm h}^2\right)\right] \left(\tilde{T_{\rm h}} \omega_{\rm c}^2-\tilde{T_{\rm c}} \omega_{\rm h}^2\right)}{\tilde{T_{\rm c}} \omega_{\rm h}^2 \left[\gamma^2 \left(\omega_{\rm c}^2+\omega_{\rm h}^2\right)+2\left(3 \omega_{\rm c}^4+10 \omega_{\rm c}^2 \omega_{\rm h}^2+3 \omega_{\rm h}^4\right)\right]-2 \tilde{T_{\rm h}} \omega_{\rm c}^2 \left[\omega_{\rm h}^2 \left(\gamma^2+7 \omega_{\rm h}^2\right)-\omega_{\rm c}^2\left(\omega_{\rm c}^2-10\omega_{\rm h}^2\right)\right]}.
\end{equation}

\begin{equation}
    Q_{\rm h}^{_{\bf L}} = 
    \frac{\gamma t \left\{2 \tilde{T_{\rm h}} \omega_{\rm c}^2 \left[\omega_{\rm h}^2 \left(\gamma^2+7 \omega_{\rm h}^2\right)-\omega_{\rm c}^2\left(\omega_{\rm c}^2-10\omega_{\rm h}^2\right)\right]-\tilde{T}_{\rm c} \omega_{\rm h}^2 \left[\gamma^2 \left(\omega_{\rm c}^2+\omega_{\rm h}^2\right)+2\left(3\omega_{\rm c}^4+10 \omega_{\rm c}^2 \omega_{\rm h}^2+3 \omega_{\rm h}^4\right)\right]\right \}}{8 \omega_{\rm c}^2 \omega_{\rm h}^2 \left[\gamma^2+8 \left(\omega_{\rm c}^2+\omega_{\rm h}^2\right)\right ]}.
\end{equation}

However, in general, the expressions of the efficiency and hot heat get more complicated. Even when  $\tau_{\rm ch }(= \tau_{\rm hc})\sim\tau_{\rm h} (= \tau_{\rm c })\ll 1$, they are already messy. 
This is why we consider the condition of $\tau_{\rm ch }(= \tau_{\rm hc})\ll t (= \tau_{\rm h} = \tau_{\rm c })\ll 1$ to show the above expressions.

\section{Quantum Refridgerator}
\label{appendix-D}

\begin{figure}[h]
 \includegraphics*[width=0.6\columnwidth]{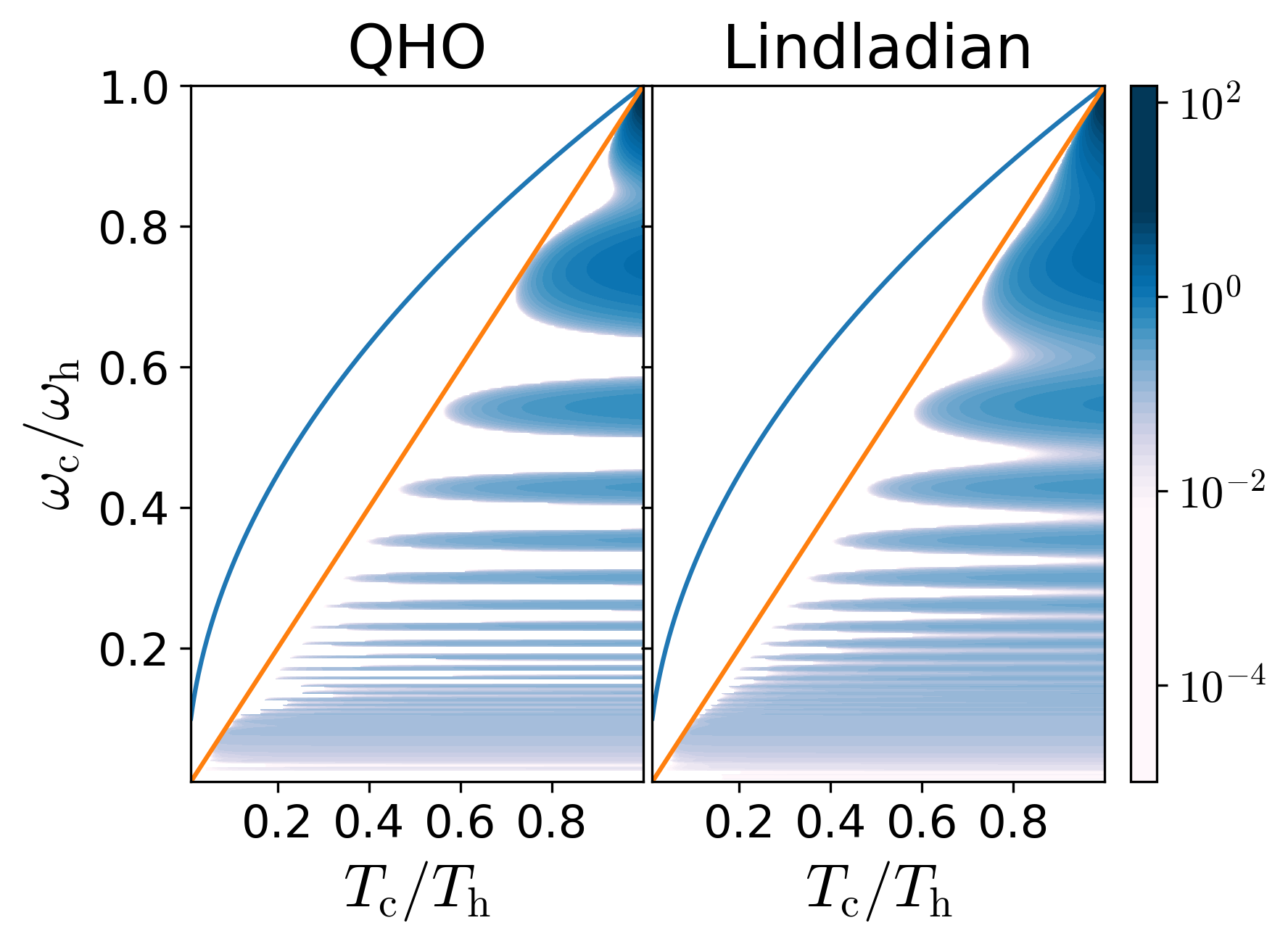}
\caption{Contour plots of cooling COP for the quantum Otto cycle in the logarithmic scale.}
\label{fig:cop}
\end{figure}

Finally, we briefly check out the cooling coefficient of performance (COP=$|Q_{\rm c}/\mathcal{W}|$) for two baths in Fig.~\ref{fig:cop}, where the region of $\omega_{\rm c}/\omega_{\rm h} < T_{\rm c}/T_{\rm h}$ is only valid. The cooling COP also exhibits wavy pattern, similar to the resonant patterns of the efficiency. The quantum Otto refridgerator was also studied by Kosloff and Rezek~\cite{Kosloff2017-QuantumOtto} as well as the quantum Otto engine.

\bibliography{ref-EL11773-LHPJ-final}

\end{document}